\documentclass[jcp,prb,twocolumn,floatfix,superscriptaddress]{revtex4-2}
\usepackage{amsfonts,amsmath,amsthm,amssymb,mathtools,bm}
\usepackage{graphicx,graphics,color,float}
\usepackage{microtype}
\usepackage{hyperref}

\begin{document}
\title{A Windowed Mean Trajectory Approximation for Condensed Phase Dynamics}
\author{Kritanjan Polley}
\email{kpolley@lbl.gov}
\affiliation{Chemical Sciences Division, Lawrence Berkeley National Laboratory, Berkeley, CA 94720, USA}
\affiliation{Department of Chemistry, University of California, Berkeley, CA 94720, USA}

\begin{abstract}
We propose a trajectory-based quasiclassical method for approximating dynamics in condensed phase systems. Building upon the previously developed Optimized Mean Trajectory (OMT) approximation that has been used to compute linear and nonlinear spectra, we borrow some ideas from filtering trajectory methods to obtain a novel semiclassical method for the dynamical propagation of density matrices. This new approximation is tested rigorously against standard multistate electronic models, spin-boson model, and models of the Fenna–Matthews–Olson complex. In all instances, the current method is significantly better or as good as many other semiclassical methods available, especially in low-temperature. All results are tested against the numerically exact Hierarchical Equations of Motion method. The new method shows excellent agreement across various parameter regimes with numerically exact results, highlighting the robustness and accuracy of our approach.
\end{abstract}

\maketitle

\section{Introduction}
Trajectory-based quasiclassical methods have been popular in computing dynamics and spectra in condensed phase systems due to computational cost comparable to a classical mechanical calculation while providing significantly accurate results.~\cite{herman94,thoss04,jasper06,kapral06,miller09} There are mostly two major categories for excited state dynamics propagation with trajectory-based methods, one based on surface hopping approaches~\cite{tully90,muller97,mac02,shenvi11,barbatti11,subotnik16} and the other based on Ehrenfest-like propagation schemes.~\cite{kab02,mannouch22,meyer79a} Surface hopping type approaches choose the force to be decided by only one potential at a given time whereas, in Ehrenfest-type methods, the force is determined by the population-averaged potential. Most mixed quantum classical approaches treat the electronic degrees of freedom quantum mechanically or semiclassically, while the nuclear degrees of freedom are evolved classically. Although ab initio methods can be easily combined with many quasi-classical approaches.~\cite{li05,park17,vindel21} Fewest-switches surface hopping~\cite{tully90} based approaches suffer from `overcoherence' problem as the stochastic jumps make these methods time-irreversible.~\cite{jain22}

The description of nonadiabatic quantum dynamics in terms of mapping Hamiltonian methods has found success in weak to moderate system-bath coupling regions. In most cases, like major semiclassical approaches, mapping Hamiltonian methods produce exactly correct result at initial time and then becomes only approximate. Quantum mechanically exact simulation of a general anharmonic Hamiltonian for a large system is still prohibitively expensive and limited to only a few degrees of freedom.~\cite{meyer90,wang03} Semiclassical approximations shine in that area as they can be applied to large and biologically relevant systems. Many approaches have been successful in this area, like the initial-value representation method~\cite{miller70,heller91}  where time propagator is usually expressed as an average over initial phase-space conditions of classical trajectories. Other approaches, such as the Forward-Backward Trajectory Solution~\cite{hsieh12,hsieh13}, the quantum classical Liouville equation (QCLE)~\cite{kim08,kelly12}, and various flavors of linearized semiclassical initial value representation~\cite{liu15,sun98,wang99,ananth07,martin07} have been tested on open quantum systems. Gao and Geva~\cite{gao20c} have shown that the accuracy of the linearized semiclassical methods can strongly depend on the choice of the zero point energy parameter, whose choice depend on the system and hence not general.  Most QCLE based approaches fail at longer times for both strong and intermediate coupling.~\cite{uken13} Recently, Mannouch and Richardson~\cite{mannouch23} proposed combining the spin-mapping approach with surface hopping dynamics called Mapping Approach to Surface Hopping (MASH) which applies to a two-state system. Runeson and Manolopoulos~\cite{runeson23} extended that method for multiple electronic degrees of freedom and tested for several exciton energy transfer models.

Optimized mean trajectory (OMT) approximation was developed for the computation of linear and nonlinear spectroscopy for condensed phase systems.~\cite{polley20b,loring22} The OMT approximation for vibronic systems is based on Meyer-Miller-Stock-Thoss (MMST) mapping~\cite{mccurdy79,meyer79a,meyer80,thoss99,thoss00} of the electronic levels and treating the nuclear degrees of freedom classically. OMT imposes specific rules on the treatment of radiation-matter interaction to mitigate the `sign problem'~\cite{shao99,kaledin03} in semiclassical mechanics. The propagation scheme in OMT is Ehrenfest-like, and the initial electronic population starts from a delta function distribution of classical action values. For the evolution of density matrix elements, OMT is equivalent to MMST mapping propagation.

The symmetrical quasi-classical (SQC) approximation is another trajectory based approximation proposed by Cotton and Miller.~\cite{cotton13a,cotton13b,cotton14,cotton19b} It makes use of the MMST mapping to obtain a classical limit of the quantum Hamiltonian and applies window functions for the initial and final trajectories to determine the dynamics of the system. The difference with Ehrenfest approach is the selection of the zero point energy parameter and application of filters to the final and initial trajectories to classify which electronic state they belong to. The judicious use of filters in SQC removes some of the drawbacks of MMST mapping and makes it an attractive candidate for application in condensed phase systems.~\cite{kananenka18,sandoval18,liang18,weight21} 

Many mapping Hamiltonian based approaches tend to break down in the regions of strong diabatic coupling, or at low temperatures, and at longer times. This motivates us to develop new semiclassical methods that does not suffer in strong coupling regions while producing reasonably accurate results at a cost comparable to a classical molecular dynamics simulation. In this article, we utilize some of the ideas from SQC methodology and combine them with MMST mapping for the dynamical propagation of condensed phase system. The new method is formulated in detail in Sec.~\ref{secModel} for a general nonadiabatic Hamiltonian. In Sec.~\ref{secResult}, we apply the current method to the spin-boson model and the FMO complex models. We demonstrate that this method agrees nicely with the numerically exact Hierarchical Equations of Motion (HEOM)~\cite{tanimura89,tanimura06,tanimura14} approach. We summarize the findings of this article in Sec.~\ref{secConclusion}.

\section{Theory}\label{secModel}
Here we consider a general nonadiabatic Hamiltonian with $F$ electronic degrees of freedom,
\begin{equation}
    \hat{H}(\bm{\hat{P}},\bm{\hat{Q}}) = \frac{\bm{\hat{P}}^2}{2m} + \sum_{j=1}^{F}\sum_{k=1}^{F} |j\rangle \langle k | \hat{H}_{jk}(\bm{\hat{Q}}), \label{eqHQM}
\end{equation}
where the first term is the nuclear kinetic energy and $\hat{H}_{jk}(\bm{\hat{Q}})$ is the diabatic electronic Hamiltonian that parametrically depends on nuclear coordinates, $\bm{\hat{Q}}$, $|j\rangle$ is the $j$th electronic state in diabatic basis, and $\bm{\hat{P}}$ are nuclear momenta.

In MMST mapping, $F$ electronic levels are mapped to a $F$-dimensional coupled singly excited harmonic oscillator following Schwinger's theory.~\cite{schwinger65} The resulting Hamiltonian has a well-defined classical limit. The Hamiltonian in Eq.~\eqref{eqHQM} with classical phase space variables is written as  
\begin{align}
    H(\bm{P}, \bm{Q}, \bm{p}, \bm{q}) & = \frac{\bm{P}^2}{2m} + \sum_{j=1}^{F} \left( \frac{1}{2}p_j^2+\frac{1}{2}q_j^2 -\gamma \right) H_{jj}(\bm{Q}) \nonumber \\
    &  + \sum_{k'=1}^F\sum_{k<k'}^F \left( q_kq_{k'} + p_kp_{k'}\right)H_{kk'}(\bm{Q}) \label{eqHclassical},
\end{align}
where $\{\bm{p}, \bm{q}\}$ are the momenta and coordinate associated with electronic degrees of freedom and $\gamma$ is the zero point energy parameter. The classical equation above can be rewritten in terms of action-angle ($\bm{n}$,$\bm{\phi}$) variables,~\cite{loring17,polley19,polley20a}
\begin{align}
    H(\bm{P}, \bm{Q}, \bm{n}, \bm{\phi}) & = \frac{\bm{P}^2}{2m} + \sum_{j=1}^{F}\sum_{k=1}^{F}  M_{jk} H_{jk}(\bm{Q}), \label{eqHwithM}
\end{align}
where
\begin{gather}
     M_{jk} = \begin{cases}
     \sqrt{(n_j+\gamma)(n_k+\gamma)}e^{i(\phi_j-\phi_k)} & j\neq k \\
     \hfill n_j \hfill & j = k 
    \end{cases}, \label{eqMjkdef}\\
    q_j =\sqrt{2(n_j+\gamma)}\cos \phi_j, \\
    p_j =-\sqrt{2(n_j+\gamma)}\sin \phi_j.
\end{gather}
The mean action values are obtained for the population elements by relating the classical action ($I_n\omega$) with energy of a single harmonic degree of freedom ($E_n=(n+1/2)\hbar \omega$). This mean action can be obtained for any elements of the density matrix analogously, $|n_L\rangle \langle n_R | \Rightarrow I_{\overline{n}}=(\overline{n}+1/2)\hbar$, with $\overline{n}=(n_L+n_R)/2$. The action values $M_{jk}$ can be taken as a semiclassical analog of the dipole moment operator. They can also be used to extract density matrix elements directly. In the original OMT approximation the zero point energy parameter, $\gamma$, was taken to be $1/2$ and the initial conditions as delta function in action variables and uniform angle variables which is exact for bare electronic systems.~\cite{polley20b} In the presence of nuclear degrees of freedom, that are treated as classical degrees of freedom and coupled to the electronic modes, OMT is an approximation as are all other trajectory based semiclassical methods.~\cite{amati23,runeson22,gao20a}

\subsection{Equations of Motion}
One can choose arbitrary $M_{jk}$ variables, with a minimum number of $F$, to propagate the dynamics of the system. We add a dummy ground state and add label 0 that is uncoupled to any other states. For a system with $F$ electronic degrees of freedom, we choose the preferred variables as $\{M_{01}, M_{02}, \dots, M_{0F}\}$. The equations of motion can be obtained from, 
\begin{equation}
    \dot{M}_{jk} = \{ M_{jk}, H(\bm{P}, \bm{Q}, \bm{n}, \bm{\phi}) \}, \label{eqMdot}
\end{equation}
where
\begin{align}
    \{M_{ab}, M_{jk} \} = &  i\delta_{jb}\left(M_{ak}+\gamma \delta_{ak}\right) \nonumber \\
     & - i\delta_{ak}\left(M_{jb}+\gamma\delta_{jb}\right).
\end{align}
As the dummy ground state is uncoupled to all other states, $n_0$ remains constant at all times. Using Eqs.~\eqref{eqHwithM} and \eqref{eqMdot}, we obtain the equations of motion following
\begin{align}
    \dot{M}_{0j} & = i\sum_{k=1}^F H_{jk}(\bm{Q})M_{0j}(t), \label{eqM0jdot} \\
    \dot{\bm{Q}} & = \frac{\partial H}{\partial \bm{P}}= \bm{P}/m, \label{eqQdot}\\
    \dot{\bm{P}} & = -\frac{\partial H}{\partial \bm{Q}} = -\sum_{k,j=1}^F \frac{\partial H_{kj}(\bm{Q})}{\partial \bm{Q}} M_{kj}(t). \label{eqPdot}
\end{align}
This particular choice of $M_{jk}$ variables produces a unitary dynamics according to Eq.~\eqref{eqM0jdot}. Following the definition in Eq.~\eqref{eqMjkdef}, any elements of the electronic density matrix can be obtained as
\begin{align}
    M_{ab}(t) & = M_{0b}(t)M_{0a}^{*}(t)/(n_0 + \gamma), \\
    M_{aa}(t) & = |M_{0a}(t)|^2/(n_0 + \gamma) - \gamma,
\end{align}
where $n_0$ is constant for a given trajectory.

\subsection{Initial Conditions}
The choice of zero point energy parameter ($\gamma$) in MMST mapping has been shown to improve the accuracy of the Ehrenfest dynamics.~\cite{amati23} An unfortunate consequence of inclusion of the zero point energy parameter is that electronic populations can become negative. Application of filters to trajectories in this context has been successful, as in the SQC approximation. The SQC approximation uses an optimum value of the zero point energy parameter for most cases that mitigates the unphysical behavior. In OMT approximation, the initial condition for the action variables are chosen from a delta function distribution,
\begin{equation}
    \rho_k(\bm{n}) = \delta(n_k-1)\prod_{j\neq k} \delta(n_j). \label{eqN0}
\end{equation}
The SQC approximation replaces the delta function for action variables with window functions of width $\gamma$. The window function is applied to both the initial and final time of the trajectory in SQC keeping the propagation is Ehrenfest type. Here we adopt the idea of replacing delta function distributions for initial action values from SQC approximation and combine it with previous OMT approximation. We do not filter the trajectories at the final time and average over all trajectories without discarding any of them. We call this new method windowed-OMT (wOMT) approximation. So Eq.~\eqref{eqN0} in OMT approximation is replaced by
\begin{align}
    \rho_k(\bm{n}) & = \delta\left(\left(\sum_{j=1}^F n_j\right)-1\right)W_k(1,\gamma) \prod_{j\neq k} W_j(0,\gamma) , \\
    W_j(s,\gamma) & = n_j \in \mathtt{rand}\_\mathtt{uniform}(s-\gamma, s+\gamma),
\end{align} 
where $2\gamma$ is the window size and $s$ is the center of the window for $j^{\mathrm{th}}$ action value.

We add an extra constraint to the window function distribution of the initial action values to ensure that at time $t=0$, the initial population adds up to 1. We choose the zero point energy parameter to be $(\sqrt{3}-1)/2 \approx 0.366 $ as in the SQC method with a square window unless otherwise mentioned. This particular value for $\gamma$ have some theoretical justification as it minimizes the zero point energy leakage while providing reasonably correct result.~\cite{cotton13a} The initial angles are sampled from a uniform distribution between 0 and $2\pi$.

\subsection{Difference with SQC, Ehrenfest, OMT}
The classical Hamiltonian in OMT approximation is from MMST mapping of the electronic variables. Ehrenfest dynamics is similar to MMST but without the zero point energy correction. The SQC approximation is built on MMST mapping but replaces the fixed initial actions with a window of width $2\gamma$ and uses $\gamma$ close to $1/3$. It also filters the final trajectory using the same filter and counts how many trajectories end up in the desired final window. SQC normalized the total population at the end of propagation to conserve probability, whereas, in the current method, the initial distribution and the unitary propagation ensure the probability is conserved.

wOMT uses similar initial conditions for action variables as in SQC, with an additional constraint that they must add up to a fixed value. For the dynamical propagation of all the model systems in this article, the total initial population is set to 1. Unlike SQC, wOMT does not filter the final trajectories, and no trajectories are discarded from the averaging procedure. 

\section{Results}\label{secResult}
We have applied the present method to the spin-boson models and the Frenkel-exciton model for energy transfer in FMO~\cite{thyrhaug18,ishizaki09} complex. Several trajectory based methods, as well as surface hopping methods, have previously been applied to these problems. At low temperatures and strong system-bath coupling regions, many of these approximate methods struggle to recover correct dynamics at long times although they remain correct initially.~\cite{tao10,kim14,thoss99,chen16,mac02} All the calculations are averaged over $10^6$ trajectories. The initial conditions for the electronic states are uncoupled with the bath modes. A simple Verlet algorithm~\cite{frenkel23book} was used for the propagation of electronic and nuclear variables as given in Eqs.~\eqref{eqM0jdot}-\eqref{eqPdot}. The timestep was chosen such that at the end of wOMT propagation, less than  $0.01\%$ trajectories had a deviation in the total population of $1\%$ or less from their initial value, otherwise the timestep was lowered and run again. All results are compared with a recently developed method, multistate mapping for MASH~\cite{lawrence24,runeson23,mannouch23}, original OMT, SQC with a square window~\cite{miller16} as used in the initial condition for wOMT,  and numerically exact benchmark HEOM method. For MASH, we have used the approach described in Ref.~\onlinecite{runeson23}.

\subsection{Application to Spin-Boson Model}
First, we consider a spin-boson model. The quantum Hamiltonian is described as
\begin{align}
    \hat{H} & = \sum_{\alpha} \left(\frac{\hat{P}_{\alpha}^2}{2m_{\alpha}} + \frac{1}{2}m_{\alpha}\omega_{\alpha}^2\hat{Q}^2_{\alpha} \right) + \Delta \big(|1\rangle \langle 2 | + \textrm{c.c.}\big)\nonumber \\
     &  + \left( \epsilon + \sum_{\alpha} c_{\alpha}\hat{Q}_{\alpha}\right) \big(|1\rangle \langle 1| - |2\rangle \langle 2|\big) ,\label{eqSBh}
\end{align}
where $\alpha$ is the index for bath modes, $\epsilon$ is the energy bias between two levels, $\Delta$ is the coupling constant between them, and $c_{\alpha}$ and $\omega_{\alpha}$ are the system-bath coupling constants and the bath frequencies, respectively.  $c_{\alpha}$ and $\omega_{\alpha}$ are sampled by discretizing an Ohmic spectral density with Lorentzian cutoﬀ at high frequency,~\cite{makri99,walters17}

\begin{align}
    J(\omega) & = \frac{\pi}{2} \sum_{\alpha}^N \frac{c_{\alpha}^2}{m_{\alpha}\omega_{\alpha}}\delta (\omega -\omega_{\alpha}) \simeq \frac{\Lambda}{2} \frac{\omega \omega_c}{\omega^2 + \omega_c^2}, \label{eqJw} \\
    \omega_{\alpha} & = \omega_c \tan \left(\frac{\alpha}{N}\mathrm{tan}^{-1} \left( \frac{\omega_N}{\omega_c} \right) \right), \label{eqwa} \\
    c_{\alpha} & = \omega_{\alpha}\sqrt{\frac{\Lambda m_{\alpha}}{2N}}, \label{eqca}
\end{align}
where $J(\omega)$ is the spectral density, $\Lambda$ is the bath reorganization energy, and $\omega_c$ is the peak frequency in spectral density. $N$ is the total number of bath modes and $\omega_N$ is the maximum oscillator frequency. We have used $\omega_N=15\omega_c$, and $N=100$. The electronic population was initialized in the first diabatic state. The initial conditions for the bath oscillators were chosen from a Wigner-transformed Boltzmann distribution,
\begin{align}
    \rho (\bm{P}, \bm{Q}) & = \prod \limits_{\alpha=1}^N \frac{\xi_{\alpha}}{\pi\hbar} \exp \left[ \frac{-2\xi_{\alpha}}{ \hbar \omega_{\alpha}}\left( \frac{P_{\alpha}^{2}}{2m_{\alpha}} +\frac{m_{\alpha}\omega_{\alpha}^2 Q_{\alpha}^2}{2} \right) \right], \label{eqWigner}
\end{align}
where $\xi_{\alpha} = \tanh (\beta \hbar \omega_{\alpha} /2)$ and $\beta=1/k_{\mathrm{B}}T$ is the inverse temperature, with $k_{\mathrm{B}}$ the Boltzmann constant and $T$ absolute temperature.

\begin{figure}
    \centering
    \includegraphics[width=\linewidth]{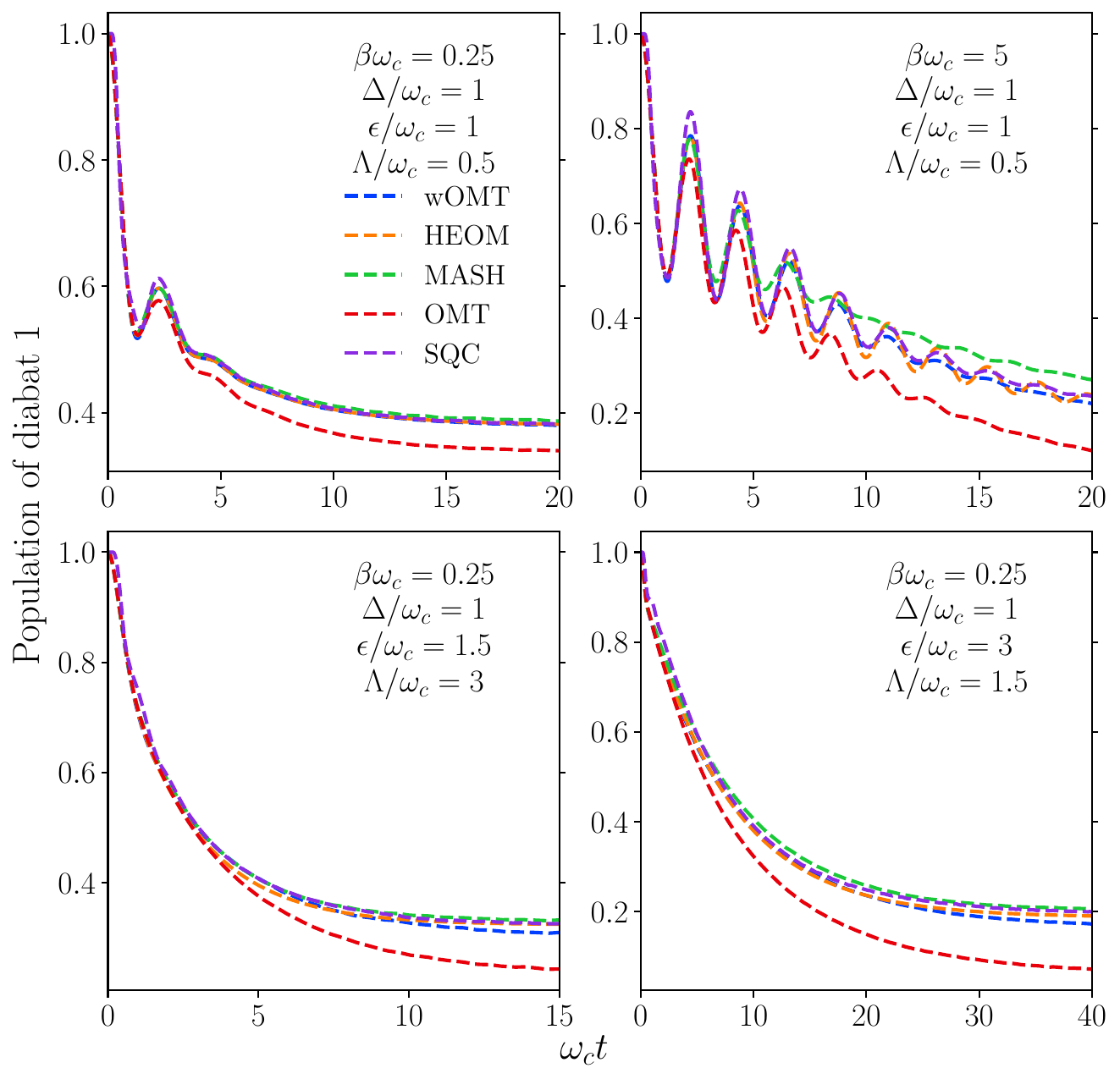}
    \caption{Population dynamics of the first diabatic state are shown above. The relevant parameters are mentioned in the panels. In the top panel, the results show coherent dynamics at high and low temperatures. In the bottom panel, the left side presents activationless electron transfer in the normal Marcus regime and on the right, electron transfer in the Marcus inverted regime. Similar parameters are sampled in Ref.~\onlinecite{mannouch23} and ~\onlinecite{runeson23}. wOMT method (blue) is compared with MASH (green), numerically exact HEOM (orange), OMT (red), and SQC (purple)  results.}
    \label{figSBpop}
\end{figure}

\begin{figure}
    \centering
    \includegraphics[width=\linewidth]{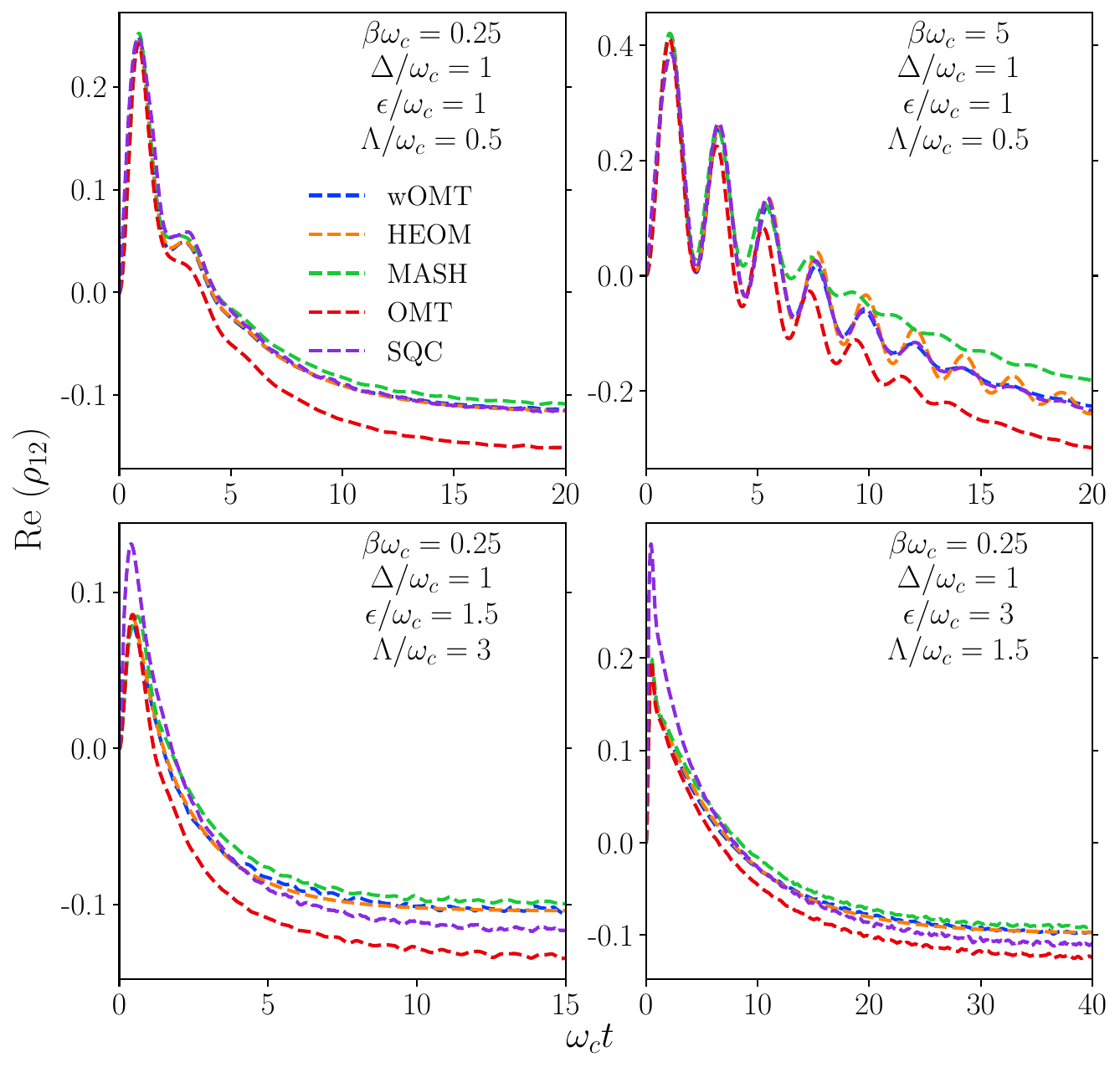}
    \caption{Real part of the 1,2 coherence dynamics with the same parameters and models as used in Fig.~\ref{figSBpop}}
    \label{figSBcoh}
\end{figure}

For population dynamics, both MASH and wOMT do reasonably well at high temperatures, although wOMT does better job at capturing the initial oscillations, as shown in the top panel in Fig.~\ref{figSBpop}. At low temperatures, one would expect a large nuclear quantum effect, but wOMT still compares better than other two approximate methods in getting the nature of the oscillations in population dynamics. For the normal Marcus regime and the inverted Marcus regime, wOMT and MASH perform nicely, compared to the benchmark HEOM results. OMT dynamics tend to agree nicely at shorter times but produce large errors at larger times irrespective of parameter choice. For coherence dynamics, similar trends are observed as for population dynamics in Fig.~\ref{figSBpop}. Interestingly, in the Marcus inverted regime, wOMT is in better agreement with coherence dynamics than population dynamics. SQC performs reasonably well, although predicts the amplitude of the initial oscillations in population dynamics to be slightly different. This is specially true for the low-temperature coherent dynamics.

Similar trends are observed in the dynamics of off-diagonal elements of the reduced electronic density matrix. The real part of the 1,2 element of the density matrix is shown in Fig.~\ref{figSBcoh}. wOMT can predict the oscillations and decay in coherence dynamics almost perfectly, except at lower temperatures where at larger times the it predicts a less oscillatory behavior.  Like in population dynamics for the spin boson models, MASH deviates from HEOM, especially in low temperature cases. Deviation from HEOM results in SQC are more prominent for off-diagonal elements, especially for normal and inverted Marcus regimes.  OMT shows a similar deviation for coherence dynamics as in the population dynamics. wOMT performs much better than OMT, highlighting the significance of the initial window sampling for the trajectories. 

\subsection{Exciton Energy Transfer}
Now, we apply the wOMT method for propagating a model for the FMO complex. Generally, the FMO complex is comprised of seven pigments, and for comparison purposes, we use two, three, and seven site/pigment models. In all cases, the approximate results are compared with HEOM calculations. SQC results with different windows for these systems have been extensively studied in Refs.~\onlinecite{cotton19a,cotton13a,cotton13b,cotton14,cotton19b}. Unless stated otherwise, 100 bath modes per site were used. The bath frequencies and coupling constants were discretized the same way following Eqs.~\eqref{eqJw}-\eqref{eqca} as in the spin boson model. The initial bath phase space variables were sampled from a Winger-transformed Boltzmann distribution, following Eq.~\eqref{eqWigner}. The model Hamiltonian is given by
\begin{align}
    \hat{H} & = \sum_j \epsilon_j |j\rangle \langle j | + \sum_{j\neq k} V_{jk} |j \rangle \langle k | \nonumber \\
    & + \sum_{j,\alpha} c_{j\alpha}\hat{Q}_{j\alpha}|j\rangle \langle j | \nonumber \\
    & + \sum_{j,\alpha} \left(\frac{\hat{P}_{j\alpha}^2}{2m_{j\alpha}} + \frac{1}{2}m_{j\alpha}\omega_{j\alpha}^2\hat{Q}^2_{j\alpha} \right), \label{eqHfmo}
\end{align}
where $V_{jk}$ are reduced electronic density matrix elements, and $\epsilon_j$ are diabatic energy levels of the electronic density matrix. Rest of the symbols carry similar meaning as in Eq.~\eqref{eqSBh}

\subsubsection{Two Site Model}
For the two-site system we choose the energy gap between two pigments $\epsilon_2 - \epsilon_1 = 100\, \mathrm{cm}^{-1}$, the coupling between them equal to $100\, \mathrm{cm}^{-1}$, and the reorganization energy, $\Lambda$ (as in Eq.~\eqref{eqJw}), to be $200\, \mathrm{cm}^{-1}$. This two state model has been extensively studied with various semiclassical approximations.~\cite{martin07,cotton16b,gao20a}

\begin{figure}
    \centering
    \includegraphics[width=\linewidth]{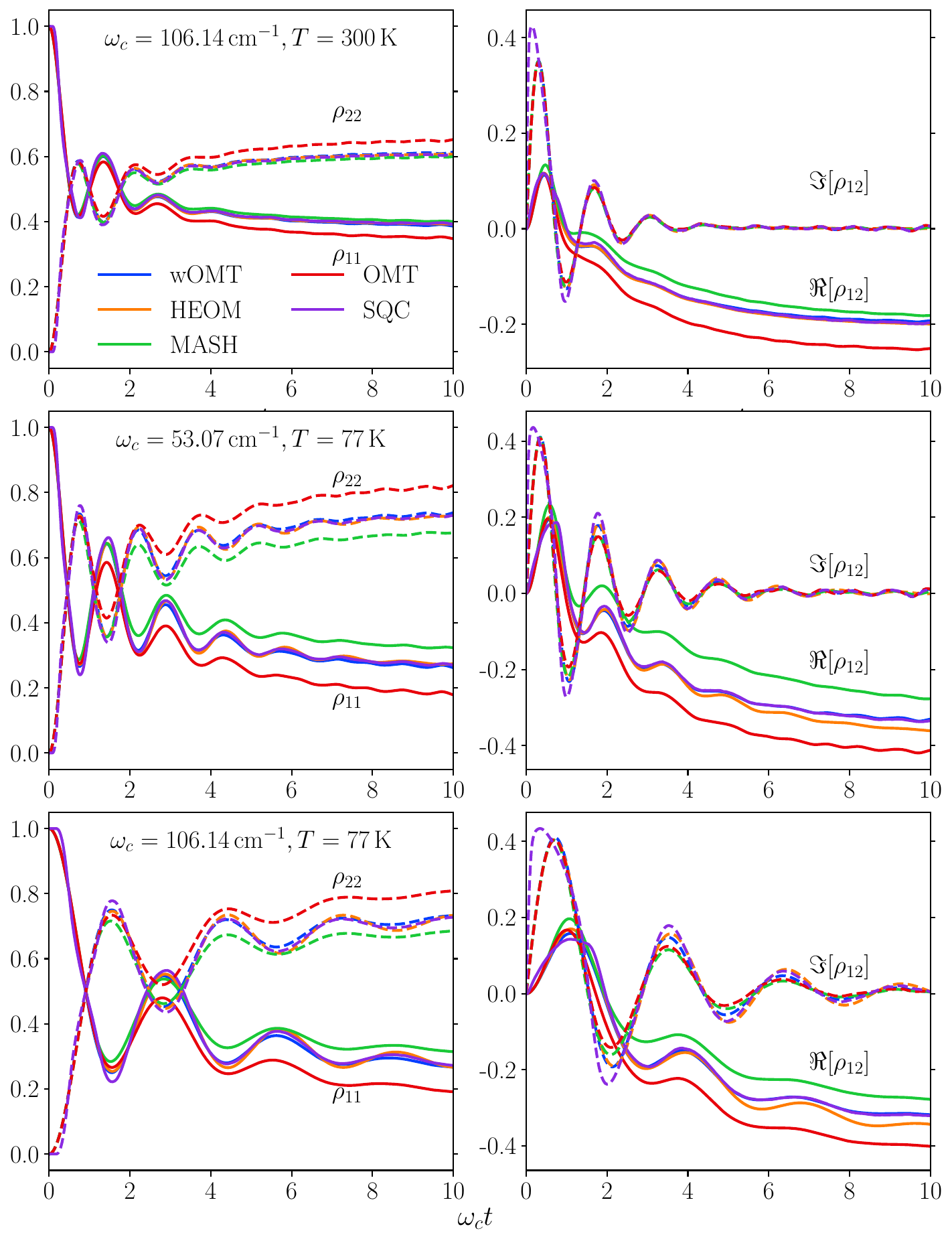}
    \caption{Population and coherence dynamics for the two-site system are displayed. The top row is for temperature $300\,\mathrm{K}$, and the remaining two are for $T=77\,\mathrm{K}$. Two values of peak bath frequency, $\omega_c=1/\tau_c$, were used. Top and middle row are for $\omega_c=53.07\, \mathrm{cm}^{-1} (\tau_c = 100\, \mathrm{fs})$ and the bottom row uses $\omega_c=106.14\, \mathrm{cm}^{-1} (\tau_c = 50\, \mathrm{fs})$. In the left column, the dashed lines are the population of the second diabat, and the solid lines are the population of the first diabat. In the right column, the dashed and solid lines are the real and imaginary parts of the 1,2 element of the electronic density matrix, respectively.}
    \label{fig2LS}
\end{figure}

Fig.~\ref{fig2LS} shows the population and coherence dynamics for different parameters for the two-site model. The left column presents the population dynamics in the two states and the right column displays the real and imaginary parts of the off-diagonal element of the electronic density matrix. wOMT performs well in most cases, with its accuracy deteriorating at longer times. It gets the initial oscillations in population and coherence dynamics almost perfectly. The efficiency of MASH deteriorates at lower temperatures and higher bath frequencies, especially at large times. Even at higher temperatures, MASH fails to get the correct value for the imaginary part of the off-diagonal elements in the electronic density matrix. In all three cases, wOMT is closer to the benchmark results than MASH. The first and the third row in Fig.~\ref{fig2LS} shows that the accuracy of wOMT decreases only slightly on going to a lower temperature. Similar to the spin-boson model, wOMT does not deviate a lot even when the temperature ($T=77 \,\mathrm{K} \equiv 53.5 \, \mathrm{cm}^{-1}$) is much smaller than peak phonon frequency. MASH predicts the trends correctly but gets the long time values incorrect, especially at lower temperatures. SQC predicts the long time values in population dynamics almost perfectly and close to wOMT. But like spin-boson models, SQC produces incorrect amplitudes for initial oscillations that becomes more evident in low-temperature regions. This is even more true for off-diagonal elements. OMT results are less accurate than wOMT for dynamics of both diagonal and off-diagonal elements, although OMT can predict the trends correctly in all cases. All semiclassical methods predict the oscillations and the subsequent decay of the imaginary part almost perfectly. For the real part of coherence dynamics, wOMT can approximate HEOM propagation best and near perfect at smaller bath peak frequency.

\subsubsection{Three and Seven Site Model}
In the site basis, the 7-site electronic Hamiltonian for the FMO model~\cite{adolphs06} is given by
\begin{equation}
    \begin{pmatrix}
    200.0 &  -87.7  & 5.5   & -5.9  & 6.7   & -13.7 & -9.9  \\
    -87.7 &  320.0  & 30.8  & 8.2   & 0.7   & 11.8  & 4.3   \\
    5.5   & 30.8    & 0.0   & -53.5 & -2.2  & -9.6  & 6.0  \\
    -5.9  & 8.2     & -53.5 & 110.0 & -70.7 & -17.0 & -63.3 \\
    6.7   & 0.7     & -2.2  & -70.7 & 270.0 &  81.1 & -1.3  \\
    -13.7 &  11.8   & -9.6  & -17.0 & 81.1  & 420.0 & 39.7 \\
    -9.9  & 4.3     & 6.0  & -63.3  & -1.3  & 39.7  & 230.0
    \end{pmatrix},
\end{equation}
where the quantities are in cm$^{-1}$ unit and, for the three-site model, the upper left $3\times 3$ submatrix is used. The energy values along the diagonal were shifted to set the minima at 0. The reorganization energy, $\Lambda$, was chosen to be 140 cm$^{-1}$ with 100 bath modes per pigment for the three site model. Other relevant parameters are mentioned in figure panels.

\begin{figure}
    \centering
    \includegraphics[width=\linewidth]{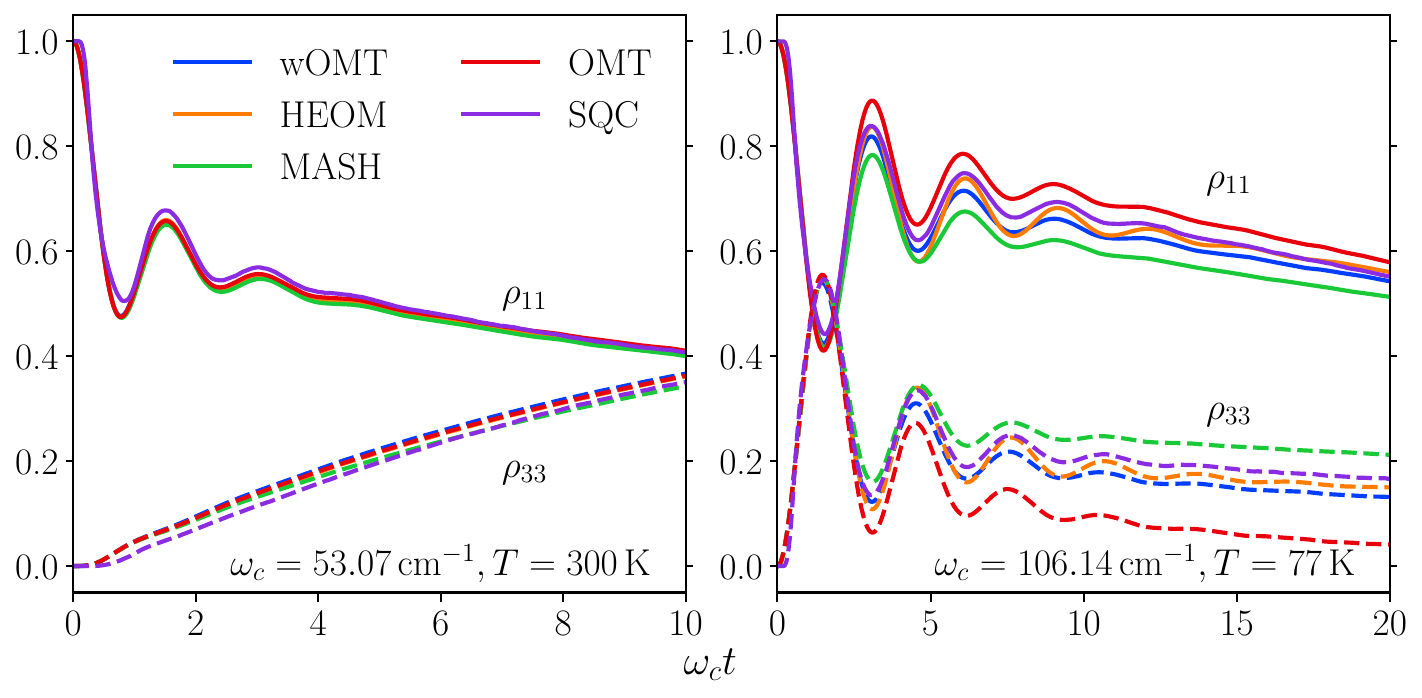}
    \caption{Population dynamics of a 3 site model of the FMO complex for wOMT method is compared with MASH, OMT, and HEOM at $300\,\mathrm{K}$ (left) and at $77\,\mathrm{K}$ (right). The electronic population was initialized in the first pigment. The populations are in site basis. Solid and dashed lines illustrate populations of the first and third diabat, respectively.}
    \label{fig3LSPop}
\end{figure}
\begin{figure}
    \centering
    \includegraphics[width=0.7\linewidth]{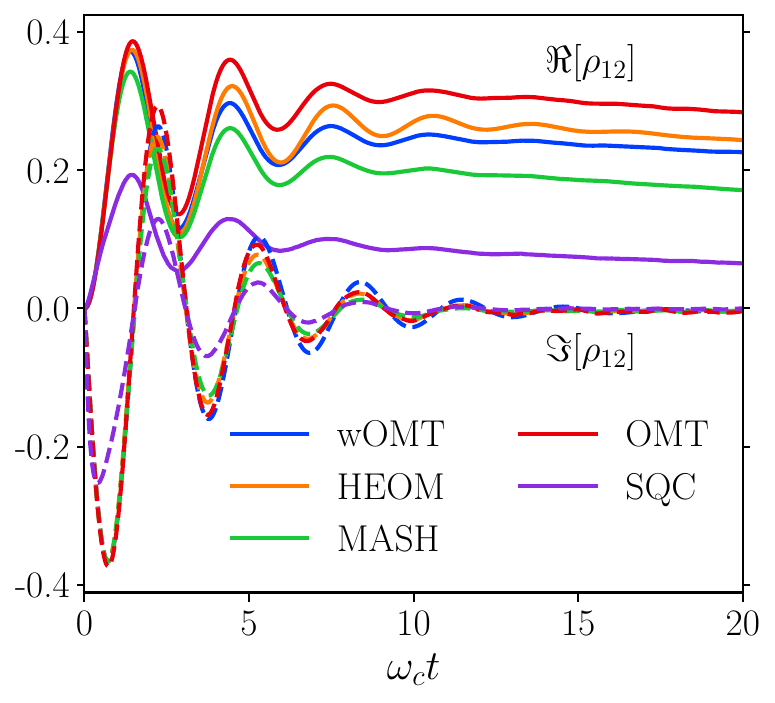}
    \caption{Dynamics of off-diagonal elements (real (solid lines) and imaginary (dashed lines) part of 1,2 element) in the three site model is demonstrated above for $T=77\,\mathrm{K}$. Same parameters and initial conditions as in the right panel of Fig.~\ref{fig3LSPop} are used.}
    \label{fig3LSCoh}
\end{figure}
\begin{center}
\begin{figure*}
    \includegraphics[width=\linewidth]{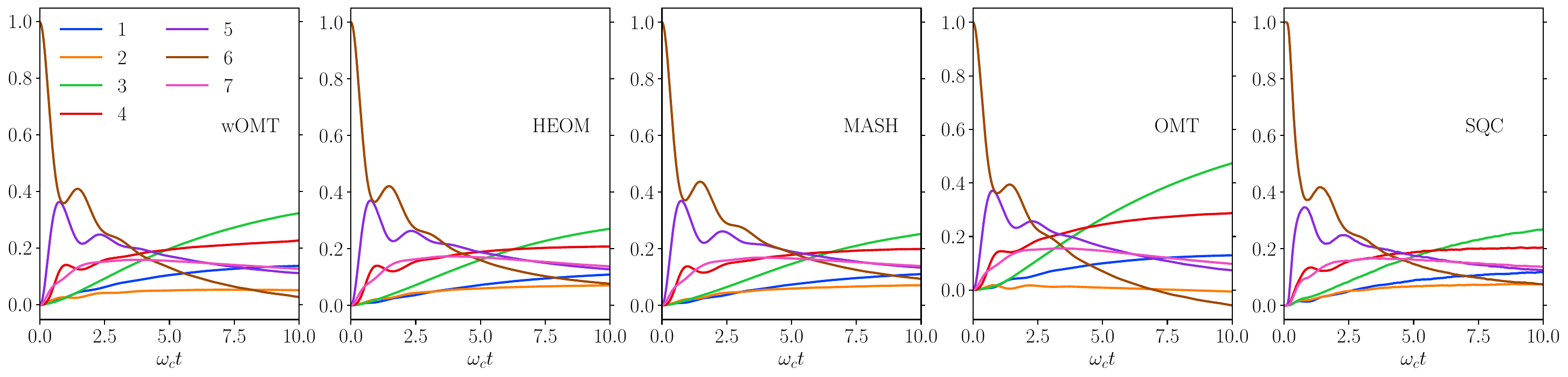}
    \caption{Population dynamics of the 7 site FMO complex model is shown above. These figures are for the same parameters as in the left panel in Fig.~\ref{fig3LSPop}, $\omega_c = 53.07 \, \mathrm{cm}^{-1}$, $T=300 \, \mathrm{K}$, and $\Lambda = 140 \, \mathrm{cm}^{-1}$. The electronic population was initiated in site 6.}
    \label{fig7LShighT}
\end{figure*}
\end{center}
\begin{center}
\begin{figure*}
    \includegraphics[width=\linewidth]{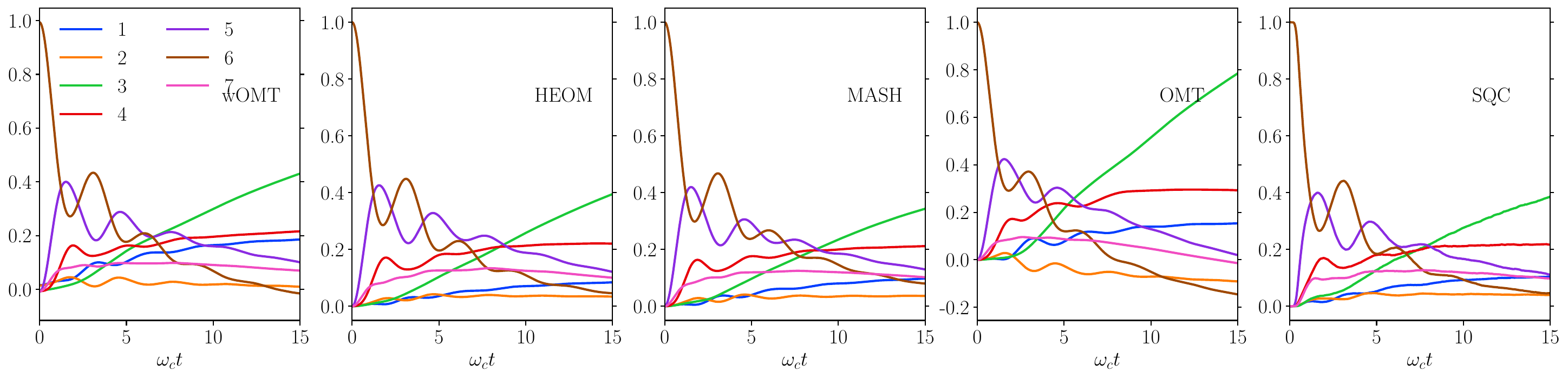}
    \caption{Population dynamics of the 7 site FMO complex model with parameters same as in the right panel in Fig.~\ref{fig3LSPop}, $\omega_c = 106.14 \, \mathrm{cm}^{-1}$, $T=77 \, \mathrm{K}$, and $\Lambda = 140 \, \mathrm{cm}^{-1}$. The electronic population was initiated in site 6.}
    \label{fig7LSlowT}
\end{figure*}
\end{center}

A decoherence-corrected surface hopping scheme was applied to the 3-site model by Sindhu and Jain.~\cite{sindhu23} Cotton and Miller tested SQC on both 3-site and seven-site models~\cite{miller16,cotton19a}.  Various other linearized semiclassical methods based on mapping Hamiltonian have also been applied to these systems.~\cite{gao20b,tao10,lee16,saller20} The population dynamics with the current method are depicted in Fig.~\ref{fig3LSPop}. The left panel shows the dynamics at 300 K while the right panel is for dynamics at 77 K. wOMT predicts the dynamics accurately at high temperature. Although its accuracy declines slightly at lower temperatures and larger characteristic bath frequency, it is still close to the benchmark HEOM result. It can reproduce the trend in oscillations right, even at later times.  The accuracy of MASH, as seen for the spin-boson model and two-site model, decreases a lot at lower temperatures. One would expect a significant nuclear quantum effect, but they seem to be small. At 300 K, OMT is as good as wOMT for population dynamics. With a larger characteristic bath frequency, OMT can predict the trend correctly but absolute values deviate from HEOM results. Similar trends are observed for the dynamics of 1,2 off-diagonal element. SQC has the largest mismatch for room temperature simulation, then it predicts the long-time dynamics nicely. Both MASH and wOMT capture the oscillations for imaginary part nicely, although they deteriorate for the real part of the off-diagonal element. MASH deviates a bit more compared to wOMT, especially at larger times. Like in two-site model, the imaginary part of the coherence dynamics, Fig.~\ref{fig3LSCoh}, is reproduced well and all three methods have similar accuracy. Again for the real part, they all predict the early dynamics nicely and deviates at later times. Although, SQC incorrectly predicts a lower long time value for the real part of the off-diagonal element.  

Results for the seven-site model are shown in Figs.~\ref{fig7LShighT} and \ref{fig7LSlowT}. For seven-site models, we have used 60 bath modes for each site. At high temperatures, in Fig.~\ref{fig7LShighT}, the initial population is in site 6. Cotton and Miller~\cite{cotton16a} have shown that time propagation of the electronic states is far more complex if the initial electronic population is located in pigment 6 than in pigment 1. MASH, although, does not get the initial oscillation of sites 5 and 6 perfectly, its long time predictions are close. OMT grows a negative population for sites 6 and 2 at larger times, and as a consequence population at sites 3 and 4 are much larger than HEOM results. SQC, like in 2- and 3-site models, predicts the correct long time dynamics for sites. wOMT follows the trends of HEOM closely at shorter times and gets the oscillatory dynamics of sites 5 and 6 almost perfectly ($\sim$ 400 fs), although wOMT predicts the long-time dynamics for sites 3 and 6 incorrectly as the population of site 6 decays more than the HEOM result. Similarly, the population of site 3 grows more, especially at longer times consistent with keeping the total population constant at all times. Other site populations are in close agreement with HEOM.

Fig.~\ref{fig7LSlowT} shows the dynamics at cryogenic temperature and a higher characteristic bath frequency. As in Fig.~\ref{fig7LShighT}, the initial population is in pigment 6.  MASH predicts a lower long-time value for state 3 and a higher long time value for state 6. wOMT predicts the early dynamics of sites 5 and 6 nicely at earlier times, especially the oscillatory behavior. It describes the rise of population in pigment 1 to be larger than the HEOM result, and the population of pigment 6 is negative at larger times, which can be a manifestation of zero point energy leakage. SQC, like the room temperature simulation, predicts the large time dynamics correctly and closest to the HEOM results among the approximate methods used here. OMT yields more negative results in the low temperature model for population dynamics. wOMT results for population dynamics of other pigments matches closely with the HEOM result.

\subsection{Limitations}
\begin{figure}
    \centering
    \includegraphics[width=\linewidth]{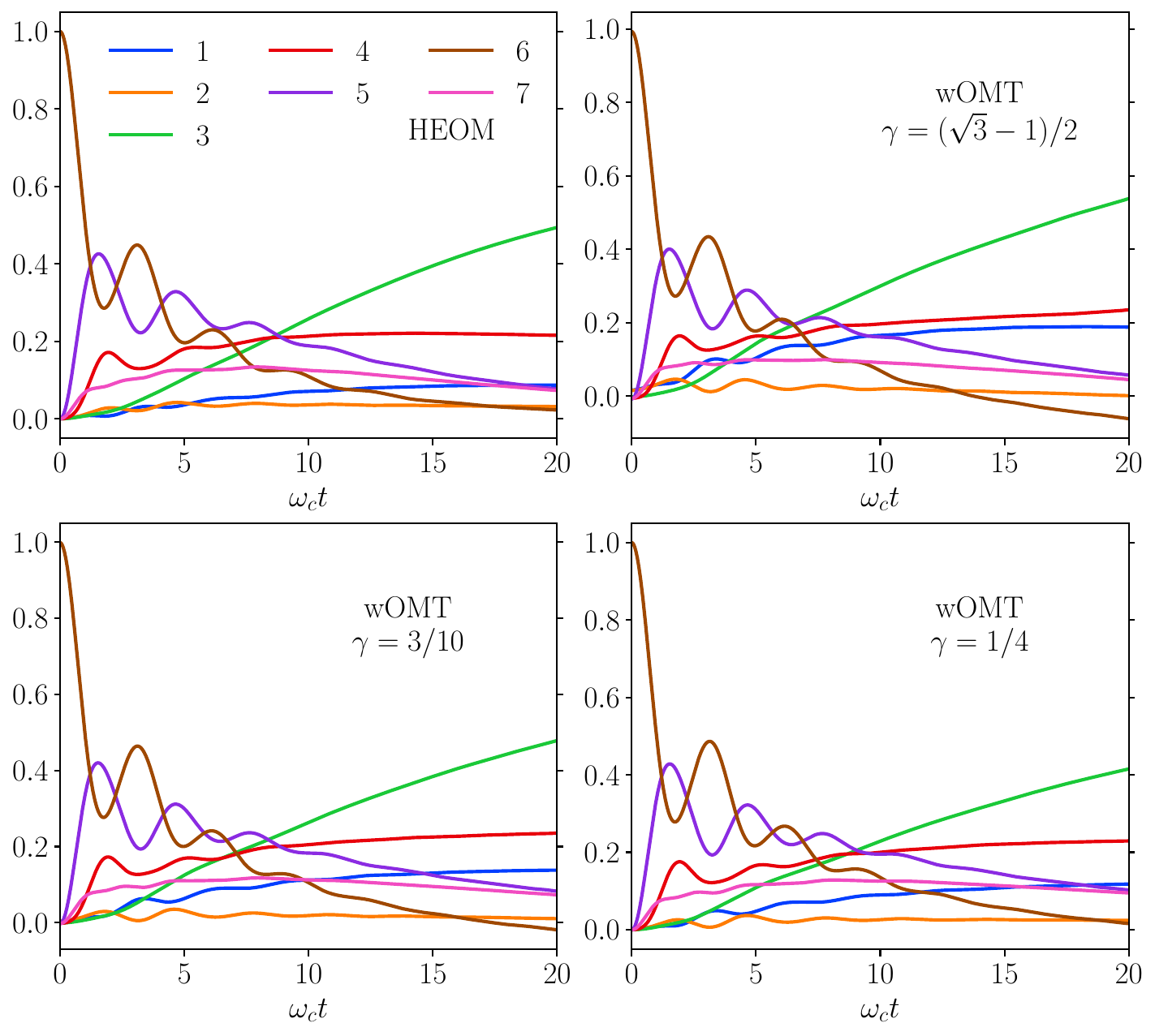}
    \caption{Comparison of the wOMT method with the HEOM method for population dynamics of the seven-site FMO complex with different zero point energy parameters. Choosing a smaller value of $\gamma$ removes the negative population but makes the earlier time population dynamics less accurate. Other parameters and initial conditions are the same as in Fig.~\ref{fig7LSlowT}}
    \label{fig7LSgamma}
\end{figure}

Being a method whose propagation scheme is Ehrenfest type, it carries similar drawbacks as observed in a normal Ehrenfest propagation. It is unable to predict the wavepacket branching in a single avoided crossing model (Tully Model 1)~\cite{runeson21}. Based on MMST mapping, wOMT naturally allows the population operators to be negative due to the presence of zero-point energy parameters. This phenomenon is generally known as the zero point energy leakage that permits populations to go negative,~\cite{muller99}, although the presence of the correction has generally improved the accuracy of MMST mapping over the Ehrenfest method. This problem is relevant for the seven-site FMO problem at low temperatures, as shown in Fig.~\ref{fig7LSlowT}. One way to treat this issue in MMST mapping is to use $\gamma$ as a variable and find an optimum value where it produces more physical as a value of $\gamma$ allows the population variables to be negative up to $-\gamma$. In the current implementation, decreasing the zero point energy removes the unphysical negative populations; but the dynamics at later times become less accurate as can be seen in Fig.~\ref{fig7LSgamma}.

For a system with many electronic levels, one can further filter the trajectories by setting individual populations that would not exceed certain thresholds. Although it improves the accuracy of wOMT slightly, the improvements for the model systems studied here are not significant.


\section{Concluding Remarks}\label{secConclusion}
In this article, we presented a straightforward yet impactful modification of the initial sampling in the original OMT approximation and applied that to the computation of population dynamics of multi-state systems. The original OMT is equivalent to MMST evolution for density matrix propagation. This simple modification demonstrated exceptional performance for spin-boson models, even in the Marcus inverted regime and low temperatures. It works admirably for models of the FMO complex and agrees nicely with numerically exact HEOM calculations. It reproduces dynamics for both population and coherences effectively. The seven-site model at low temperatures, does predict negative population at large times, but it captures the early time dynamics effectively. All the calculations presented here employed a diabatic Hamiltonian. However, as the propagation scheme is Ehrenfest-like, it can also be used in adiabatic representation. For a detailed description of the adiabatic version of Meyer-Miller mapping, one can follow Ref.~\onlinecite{cotton17}. 

As pointed out by Lawrence \textit{et al.}~\cite{lawrence24}, the version of MASH used in this article is likely to suffer when the nuclei are more strongly coupled to the electronic coordinates. We did find that to be ture, especially at low temperatures. SQC predicts the long time dynamics well, although it gets the amplitudes of the early oscillations slightly wrong. The deviation in SQC method for off-diagonal elements can be understood from the fact that windowing procedure allows a poor averaging for the highly oscillatory phase factor which is absent for population terms. Removing the final window improves the wOMT averaging for the coherence terms. 

Finally, we comment on the convergence of these methods with the number of trajectories used. Although the primary aim of this paper was to evaluate the accuracy of this current approach, and we used 10$^6$ trajectories to generate the figures here, a rough qualitatively similar result can be obtained with approximately $10^4 - 10^5$ trajectories. The results obtained with 10$^6$ trajectories exhibited trends consistent with the ones using 10$^4$ trajectories.  The averaging gets better with more trajectories, leading to better-converged curves. This is similar to other trajectory based methods applicable to the problem studied here.~\cite{cotton16a,kananenka18} Computational costs for wOMT and OMT are similar as same equations of motion are propagated in both cases, just with different initial conditions. The present implementation of MASH is slightly costlier than the other two as it is performed in adiabatic basis and switch between adiabatic and diabatic potential on every time step makes it slightly more expensive. Although wOMT method simulated in an adiabatic basis will have similar cost.

A simple modification in the initial sampling of electronic states vastly improves the accuracy of the OMT method. It can certainly be extended for the computation of linear and nonlinear spectra, which will be an interesting future study. Also, inspired by the development of MASH methods~\cite{lawrence24,runeson23,mannouch23}, a surface hopping propagation scheme implementation in the current method is likely to overcome the limitations it inherits due to Ehrenfest dynamics like failure with a single avoided crossing model. It is interesting that nuclear quantum effects were not very large, even at low temperatures. In cases where the nuclear quantum effects are significant, one way to incorporate that would be to combine ring polymer molecular dynamics~\cite{habershon13,ananth22} scheme with the current method as it was done with the surface hopping approach.~\cite{shushkov12,shakib17} Continued inspection and refinement of these techniques will provide us with important tools for exploring and understanding the condensed phase dynamics.

\section*{Acknowledgement}
This work was supported by the Condensed Phase and Interfacial Molecular Science Program (CPIMS) of the U.S. Department of Energy under Contract No. DE-AC02-05CH11231. The author thanks David Limmer, Roger Loring, Eric Heller, and Amr Dodin for providing useful comments on the manuscript.

\section*{Data Availability}
All the codes and data to reproduce the figures in this paper are available at \href{https://github.com/kritanjan-polley/wOMT.git}{https://github.com/kritanjan-polley/wOMT.git}



\bibliography{reference}

\end{document}